\documentclass{article}
\usepackage{icrctc07}
\pdfoutput=1
\title{Feasibility of acoustic neutrino detection in ice: 
\newline 
Design and performance of the South Pole Acoustic Test Setup (SPATS)}
\shorttitle{South Pole Acoustic Test Setup}
\authors{ S. B\"{o}ser$^1$, C. Bohm$^2$, F. Descamps$^{3}$,
  J. Fischer$^1$, A. Hallgren$^4$, R. Heller$^1$, S.~ Hundertmark$^2$,
  K. Krieger$^1$, R. Nahnhauer$^1$, M. Pohl$^1$, P. B. Price$^5$,
  K.-H. Sulanke$^1$, D.~ Tosi$^1$ and J. Vandenbroucke$^5$}

\shortauthors{F. Descamps et al.}
\afiliations{
$^1$ DESY, Platanenallee 6, D-15738 Zeuthen, Germany\\$^2$ University
  of Stockholm, Fysikum, SE-10691 Stockholm, Sweden\\$^3$ University of
  Ghent, Department of Subatomic and Radiation Physics, Belgium\\$^4$
  University Uppsala, Department of Nuclear and Particle Physics, Box 535,
  SE-751 21 Uppsala, Sweden\\$^5$ Dept. of Physics, University of
  California, Berkeley, CA 94720, USA
}
\email{Freija.Descamps@Ugent.be}

\abstract{
The South Pole Acoustic Test Setup (SPATS) has been built to evaluate
the acoustic characteristics of the South Pole ice in the 10 to 100 kHz
frequency range so that the feasibility and specific design of an
acoustic neutrino detection array at South Pole can be
evaluated. SPATS consists of three vertical strings that were
deployed in the upper 400 meters of the South Pole ice cap in January
2007, using the upper part of IceCube holes. The strings form a
triangular array with the longest baseline 421 meters. Each of them
has 7 stages with one transmitter and one sensor module. Both are
equipped with piezoelectric ceramic elements in order to produce or
detect sound. Analog signals are brought to the surface on electric
cables where they are digitized by a PC-based data acquisition
system. The data from all three strings are collected on a
master-PC in a central facility, from which they are sent to the
northern hemisphere via a satellite link or locally stored on tape. A
technical overview of the SPATS detector and its performance is presented.

}

\begin{document}
\maketitle
%Begin the section.
\section{Motivation}
The effective volume needed for the detection of the predicted small
cosmogenic neutrino flux and the study of its angular distribution is orders of magnitude larger than
the instrumented volumes of the Cherenkov neutrino
telescopes~\cite{Achterberg:2006md,Aguilar:2006rm} currently under construction. New
detection methods that are sensitive to the radio and acoustic
signatures of a UHE neutrino interaction would allow a more sparse
instrumentation and therefore a larger sensitive volume at reasonable
cost. The feasibility and specific design of an acoustic array as part
of a hybrid opical/radio/acoustic neutrino detector, as was suggested
and simulated in \cite{hybrid}, depend on the acoustic properties of
the South Pole ice in the concerned frequency region (10 to 100 kHz).
The South Pole Acoustic Test Setup (SPATS) has been built to evaluate the
attenuation length, speed of sound, background noise level and
transient rates and was deployed in the 06/07 polar season.

\section{SPATS layout and in-ice components}
The South Pole Acoustic Test Setup consists of three vertical strings
that were deployed in the upper 400 meters of selected IceCube holes to form 
a triangular array, with inter-string
distances of 125, 302 and 421 meters. 
Each string has 7 acoustic stages.
The upper part of the ice has a
larger density gradient and therefore a larger variation of the acoustic properties is expected. Because of this
the distance between
stages increases with depth: they are positioned at 80, 100, 140, 190, 250, 320 and 400 m depth.
Figure 1 shows a schematic of the SPATS array and its in-ice and on-ice components.
\begin{figure*}
\begin{center}
\includegraphics [height=0.60\textwidth,width=0.80\textwidth]{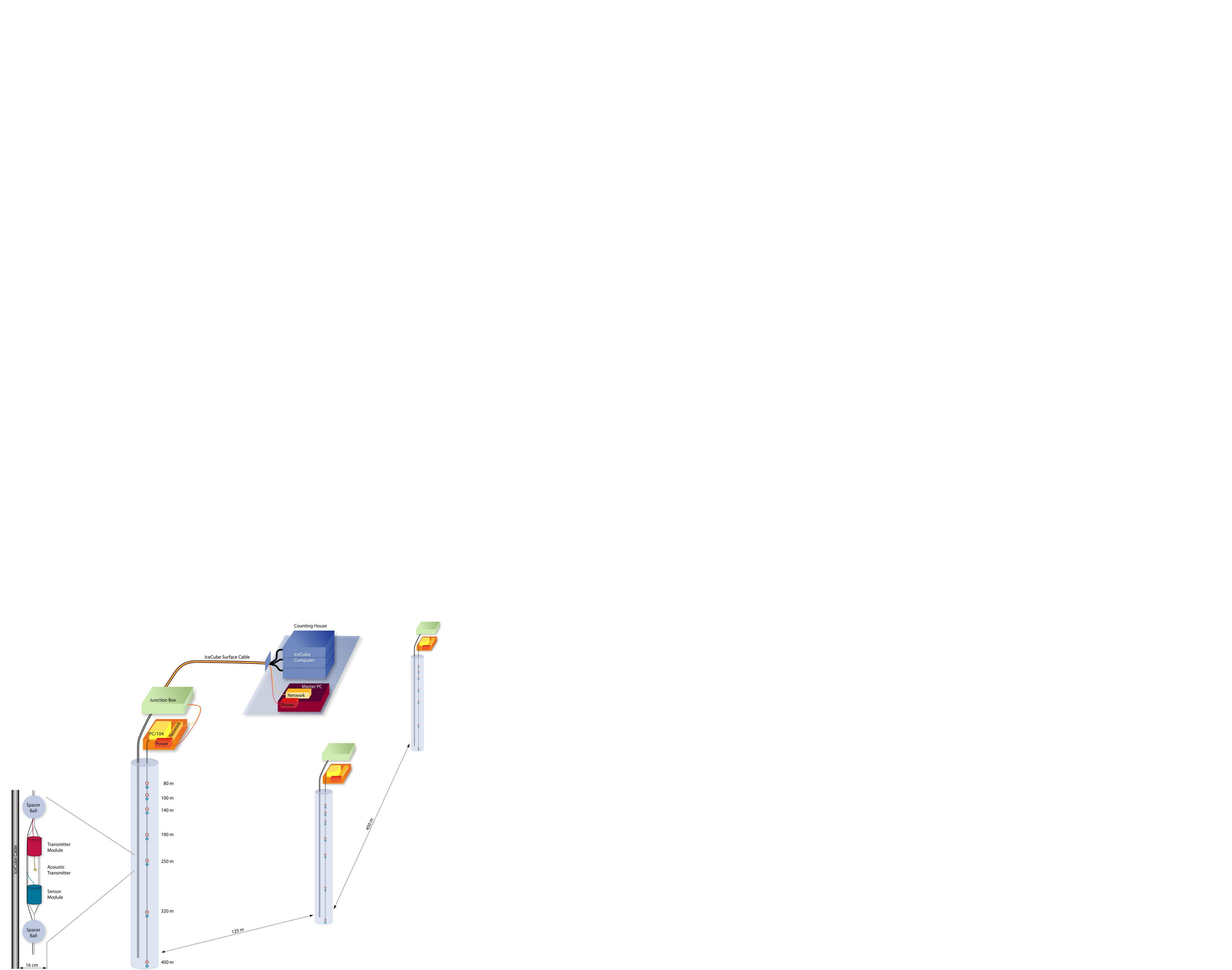}
\end{center}
\caption{Schematic of the SPATS array, with the three strings each with 7 acoustic stages. 
}\label{fig1}
\end{figure*}

An acoustic stage consists of a transmitter module and a sensor module. All the
electronic circuits are located in steel pressure housings with a diameter of
 10.2 cm. Figure 1 also shows a shematic drawing of an acoustic stage.
In total 25 stages were produced and tested.

The sensor module has three channels to ensure good angular coverage. Each
channel consists of a piezo-ceramic element that is closely attached
to a 3-stage amplifier with an amplification on the order of
10$^{4}$. A voltage regulation board provides $\pm$5 V and a virtual ground.
The complex mechanical design of the sensor module induces a multitude of resonances over 
the full frequency range. In order to obtain a feeling of the variation of the
response between the different sensors, a calibration in water was
performed and the equivalent self noise spectrum was extracted for each sensor channel.
Figure 2 shows the range of equivalent self noise for all SPATS sensors as a function of frequency. 
The pressure amplitude needed to overcome
the sensor self noise ranges from 11 to 83 mPa. This is overall better performance than the commercial 
hydrophone that was used as reference (150 mPa).

\begin{figure}
\begin{center}
\includegraphics [height=0.43\textwidth,width=0.52\textwidth]{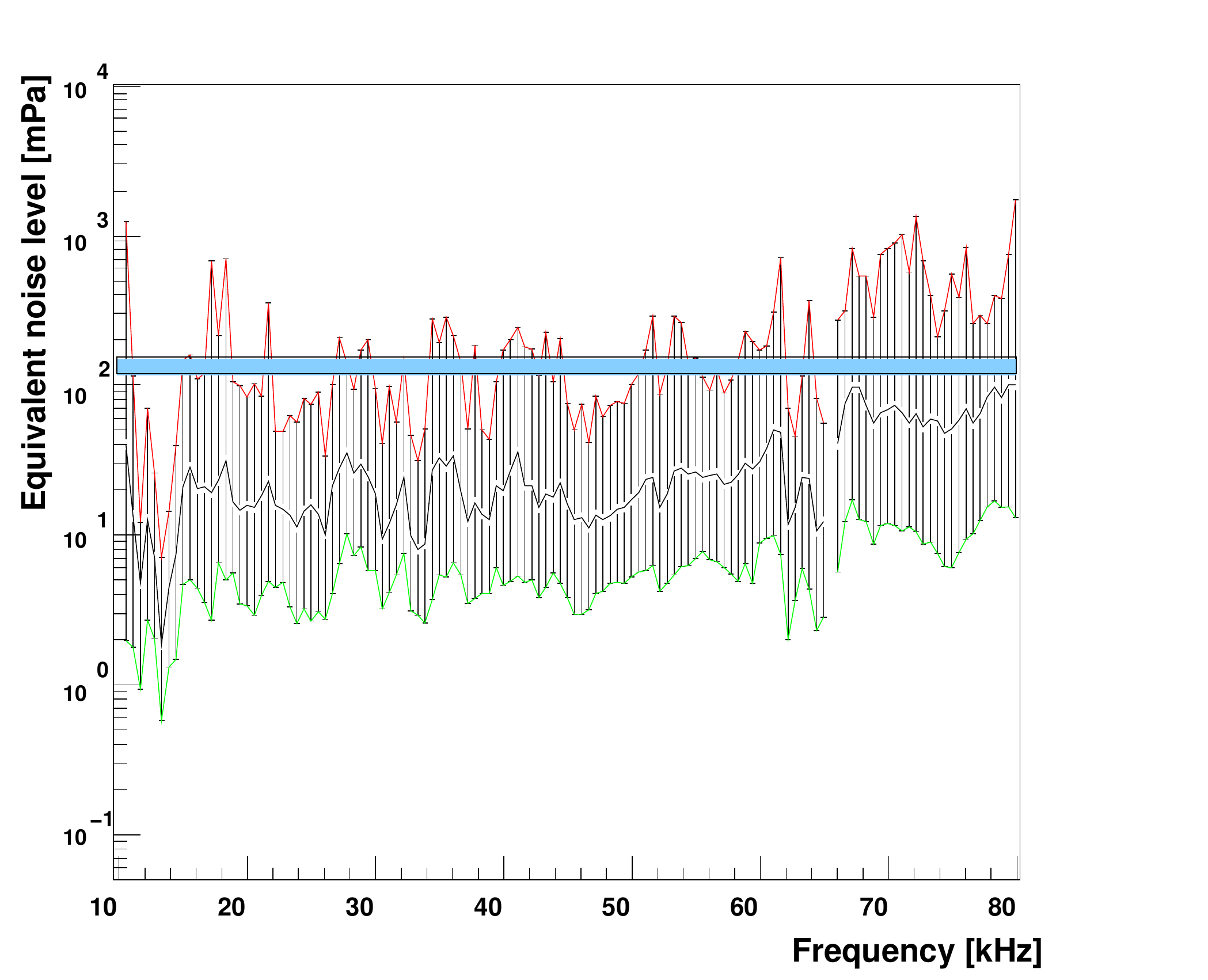}
\end{center}
\caption{Equivalent noise level of all SPATS sensor module channels; the bars indicate 
the range over which all the sensors are spread. The curved black line corresponds to the mean value. 
The hydrophone level is indicated by a thick band.}\label{fig2}
\end{figure}

The transmitter module contains an LC circuit that provides
sinusoidal half-wave high voltage pulses with a FWHM of about 10 $\mu$s and a maximum of around 1 kV. These are
triggered by TTL signals and then sent to the transmitter: a ring-shaped
piezo-ceramic element that is cast in epoxy for electrical insulation
and positioned $\sim$13 cm below the steel housing. 
Azimuthally isotropic emission is the motivation for the use of ring shaped piezo-ceramics. The actual
emission directivity of such an element was measured in azimuthal and polar directions. It was found that 
an uncertainty of 40$\%$ can be expected due to a unknown azimuthal transmitter orientation. 
 More details on the water sensor and transmitter calibration can be found in~\cite{arena}.
Each of the stages is connected to the surface by 2
shielded cables of 4 twisted pairs.
\section{Data acquistion and system tests}
The analog signals that arrive through the in-ice cable at the surface
are digitized directly at the hole location in order to allow readout of the 
many channels on a small number of surface cable pairs. To this end an acoustic 
junction box containing electronic
components was installed in the surface snow. The robust aluminium box is split in two compartments; one
holds the sockets to connect the cables and the other contains a
printed circuit board and a low-power industrial PC. This {\it
  string-PC} is a PC/104 system that is controlled by a CPU module with a 600 MHz
processor and 512~Mb RAM. Three fast ADC boards allow for all three
channels of one sensor module to be read out simultaneously at 1.25 MHz
sampling frequency up to all channels of one string at $\sim$179 kHz. 
Together with one slow ADC/DAC board with a sampling frequency of
500~ kHz, the transmitters can be controlled and the temperature and
pressure sensors read out. A relay board switches the power for each sensor and transmitter 
on and off separately so that the average power consumption per string 
stays rather low ($\sim$35 W). All these components were tested at the expected low temperatures (around -55 $^{ o}$C) and several cold boot cycles of the system were successfully performed. 

 The acoustic junction box of each string is connected to a
{\it master-PC} located in the IceCube Laboratory  by two four-wire
cables of the IceCube surface cable. 2 pairs are used to provide 96 V to the DC/DC converters that
generate the required voltages for the string. Communication is provided by a  
2~ Mbps symmetric DSL connection and a  GPS-based IRIG-B time coding signal
guarantees synchronisation between the different strings and allows 10 microsecond absolute time-stamping. 
The communication and timing signal as well as the supplied voltage are routed through a PCI control
 board in the master-PC that provides current and voltage control and monitoring.

The {\it master-PC} can be accessed through the South Pole network via a satellite link, 
which allows remote control of the system. A selection of the generated data is currenltly 
transferred north for analysis at a rate of 150 Mb/day. The rest are written to magnetic tape 
and will be transported north in the next austral summer.
 
A SPATS system test was performed in lake Tornetr\"{a}sk in northern Sweden in April 2006. The lake was at that
time still covered with 90 cm of ice, allowing easy access and deployment.
The transmitter range in water was measured and the performance of the sensors was compared 
to that of a commercial hydrophone.
\begin{figure}[h]
\begin{center}
\includegraphics [width=0.50\textwidth]{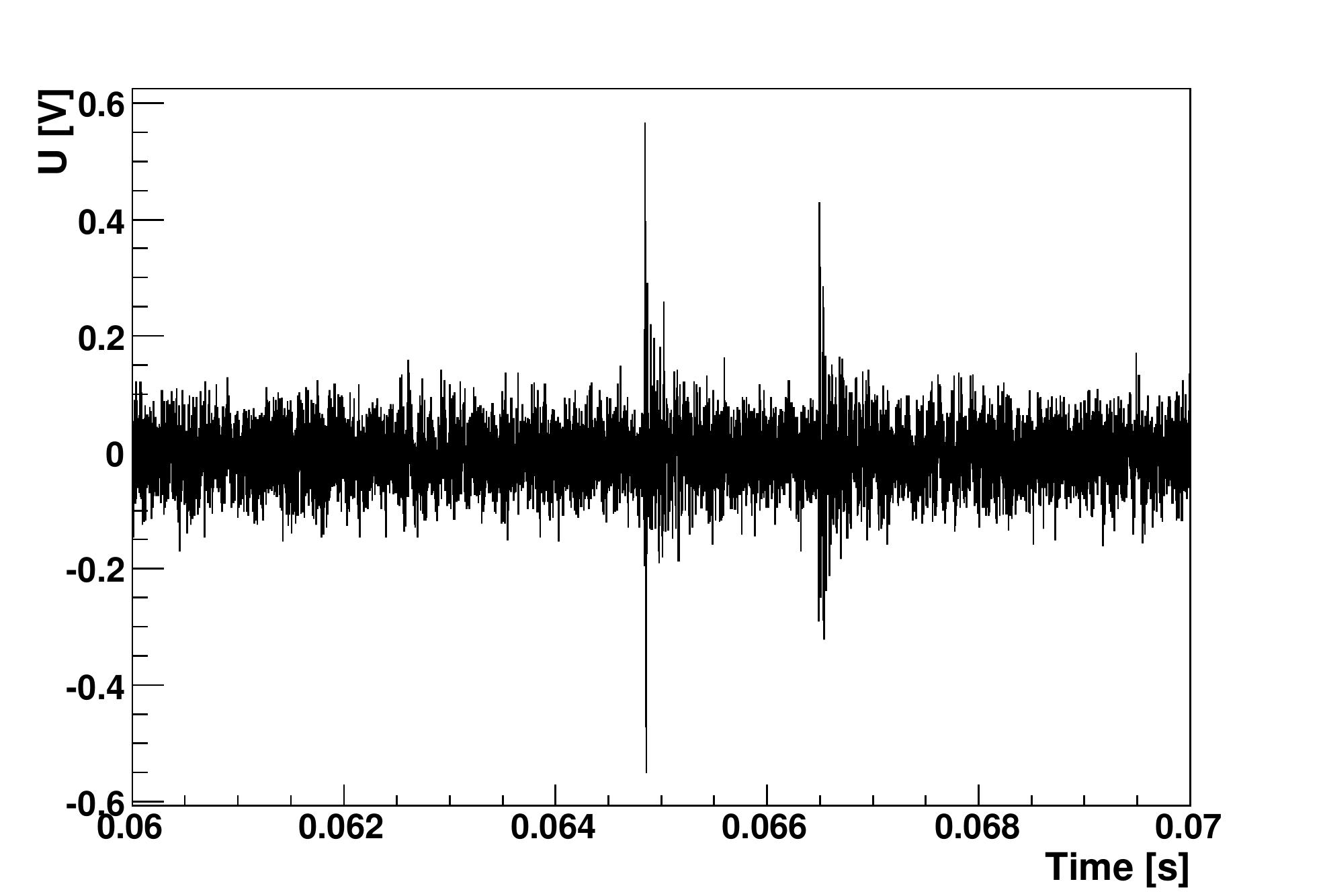}
\end{center}
\caption{The transmitter signal as seen by a SPATS sensor at 800 m distance in water.}\label{fig3}
\end{figure}
The maximum transmitter to sensor distance of 800 m was determined by the available cable length.
Figure 3 shows 
the recorded pulse with a signal to noise ratio of $\sim$10.  
There is a second pulse, reflected off the surface and delayed by $\sim$1 ms.
These data show that the transmitter range in water is larger than 800 m. 
As a rule, the signals as recorded by the SPATS sensor were much stronger than those of the commercial
hydrophone. In fact, the hydrophone was incapable of detecting a transmitter signal above noise at 400 m 
distance.

The complete set of stages was tested at South Pole before deployment in order 
to select the best 21. All selected stages 
were established to work within normal parameters. 
\section{Commissioning and current status}
The three strings were installed in January 2007 and each was commissioned within 24 hours 
after deployment while the stages were still in water. 
By recording transmitter pulses within the  same string, it was established that all stages were operational.
Figure 4 shows a commissioning event: a transmitter pulse from a stage at 100 m depth was recorded by a sensor channel of a stage at 80 ~m depth on the same string. 
\begin{figure}[h]
\begin{center}
%\fbox{\hbox{\vbox{\hsize=50mm \hfill \vspace{50mm}}}}
\includegraphics [width=0.45\textwidth]{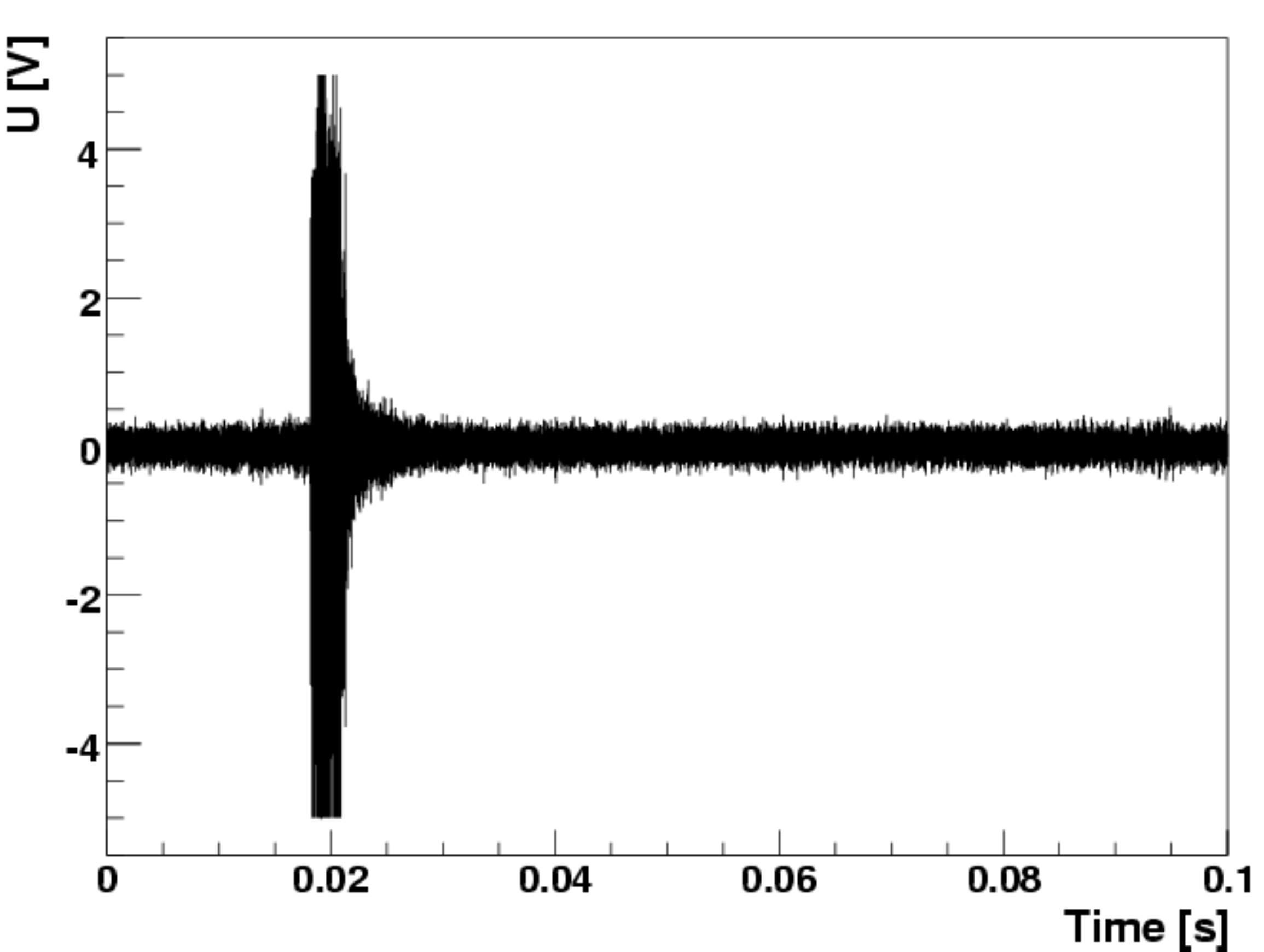}
\end{center}
\caption{An intra-string commissioning run in water. The time axis has an arbitrary offset.}\label{fig4}
\end{figure}
53 out of 63 sensor channels and all transmitters are working in ice within normal parameters. Out of the 
channels that are experiencing problems, three failed shortly after deployment and another three during freeze-in. 
The system has 
been running almost continuously since the end of January 2007, where the interruptions were solely due to
outside influences such as power outages. The PC/104 system has proven to be robust, surviving all power outages and 
subsequent cold reboots. The current status of the data analysis can be found in~\cite{justin}.
\section{Conclusions and outlook}
The South Pole Acoustic Test Setup has been succesfully deployed in the 06/07 austral summer after extensive 
testing of all components. All 21 deployed acoustic stages were calibrated in water. The data are of good 
quality.

A fourth acoustic string is being prepared for deployment in the 07/08 austral summer.
3 SPATS-like acoustic stages will be deployed to a maximum depth of 500 m.
Other acoustic devices are currently under construction; the goal is to deploy these second generation devices
at the other available breakouts of the fourth string. Continued R\&D efforts work towards a possible installation 
of a 100 km$^{2}$ scale acoustic array in the South Pole ice as part of a hybrid EeV neutrino detector. 
\section{Acknowledgements}
The deployment and success of the South Pole Acoustic Test Setup would
not have been possible without the support of the IceCube
Collaboration. 
%This is the reference to .bib file (Whitout .bib!)
%\bibliography{libros}
%This in the bibtex style, is ok.
\bibliographystyle{plain}

\end{document}